\documentclass[a4paper]{jpconf}
\usepackage{graphicx}

\usepackage{citesort}
\begin{document}
\nocite{*}
\title{Measurement of Heavy Quark cross-sections at CDF}

\author{Alberto Annovi on behalf of the CDF collaboration}

\address{Laboratori Nazionali di Frascati, Via E. Fermi 40, I-00044 Frascati (Rome), Italy}

\ead{alberto.annovi@lnf.infn.it}

\begin{abstract}
The measurement of heavy quark cross-sections provides important tests of the QCD theory. This paper reviews recent measurements of single b-quark and correlated b-quark cross-sections at CDF. Two new measurements of the single b-quark production at CDF agree with the first result from CDF~Run~II. This clarifies the experimental situation and confirms the recent agreement of theoretical prediction with data. A new measurement of the correlated $b\bar{b}$ cross-section with dimuon events at CDF is presented. It agrees with theory and it does not confirm the anomalously large $b\bar{b}$ cross-section seen in Run~I by CDF and D${\not\! {\rm O}}$ in dimuon events.
\end{abstract}

\section{Single b-quark production at the Tevatron}
Historically the b-quark cross-sections measured at the Tevatron during Run~I have been higher than expected from NLO QCD predictions.
This fact fostered continuous efforts to improve both the measurements and the theoretical predictions.
A first compatible calculation was delivered in 2002~\cite{Cacciari:2002pa} and updated in ref.~\cite{Cacciari:2003uh}.
Where the increase in predicted cross-section is mostly due to improved experimental inputs: PDF, $\alpha_s$, and especially fragmentation functions tuned for 
calculations with logarithms resummation at next-to-leading accuracy (NLL)~\cite{Cacciari:2002pa}.
These new results, while going in the right direction of a higher prediction, are almost compatible but still lower than the lastest CDF~Run~I result~\cite{Acosta:2001rz}.
The first measurement of the b-quark cross-section with CDF~Run~II data uses $p\bar{p}\rightarrow J/\psi + X$ events~\cite{Acosta:2004yw}.
This result is lower than the latest Run~I result and in perfect agreement with the predictions~\cite{Cacciari:2003uh}.

This would be the end of the story, but the experimental picture is not fully clear as noted in ref.~\cite{Happacher:2005gx}. In fact, the experimental measurements are not fully consistent among themselves.
Before claiming the solution of the long standing problem of b-quark production at the Tevatron, the experimental situation must be clarified.
With this purpose CDF provided two additional measurements 
that use different decay modes. 

\section{Recent measurements of single b-quark cross-section at CDF}
\begin{figure}[h]
\begin{minipage}[t]{17pc}
\includegraphics[width=15pc]{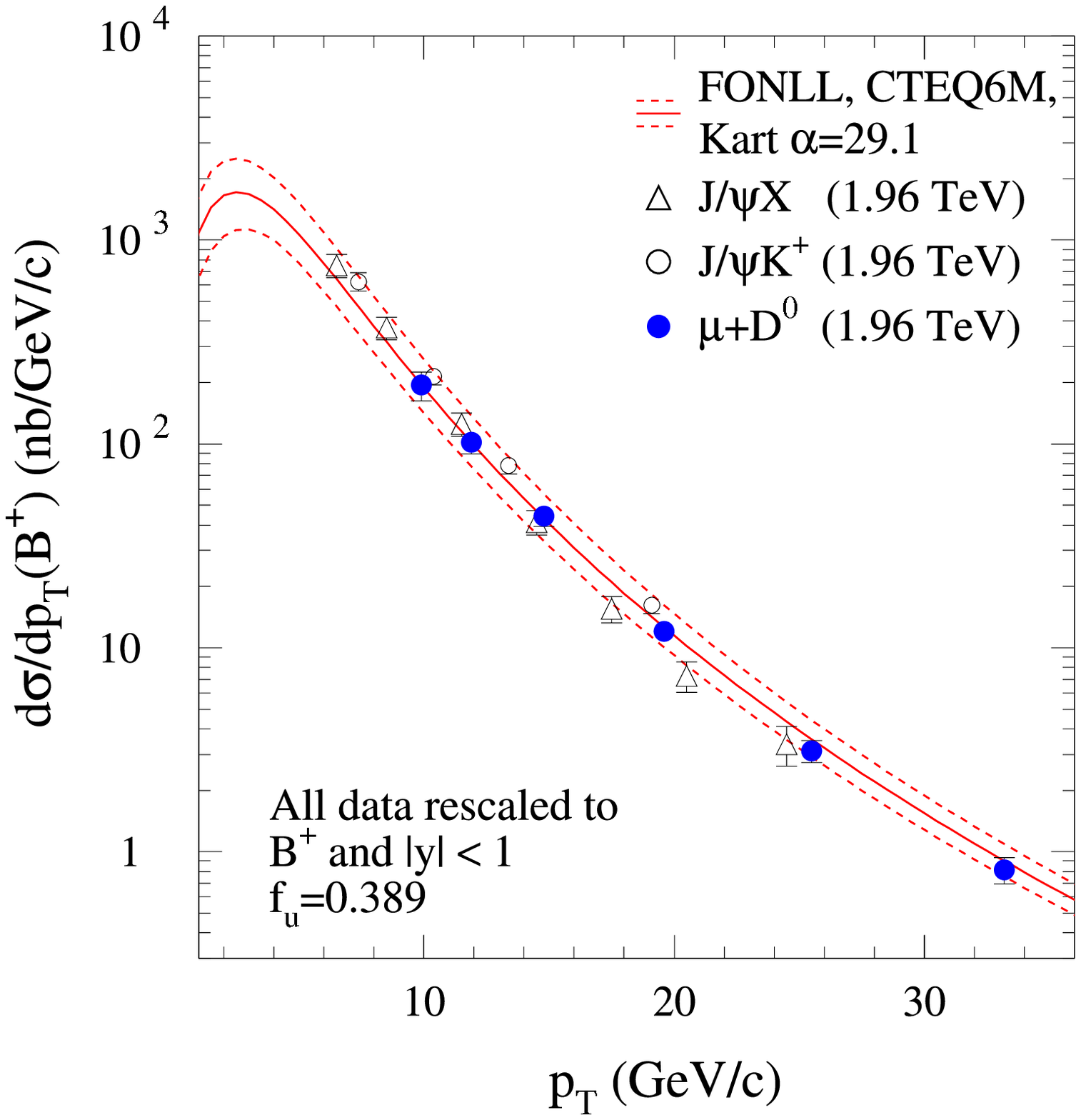}
\caption{\label{fig:comp1}CDF~Run~II measurements of b-quark production compared to FONLL.}
\end{minipage}
\hspace{2pc}%
\begin{minipage}[t]{18pc}
\hspace{-3pc}
\includegraphics[width=22pc]{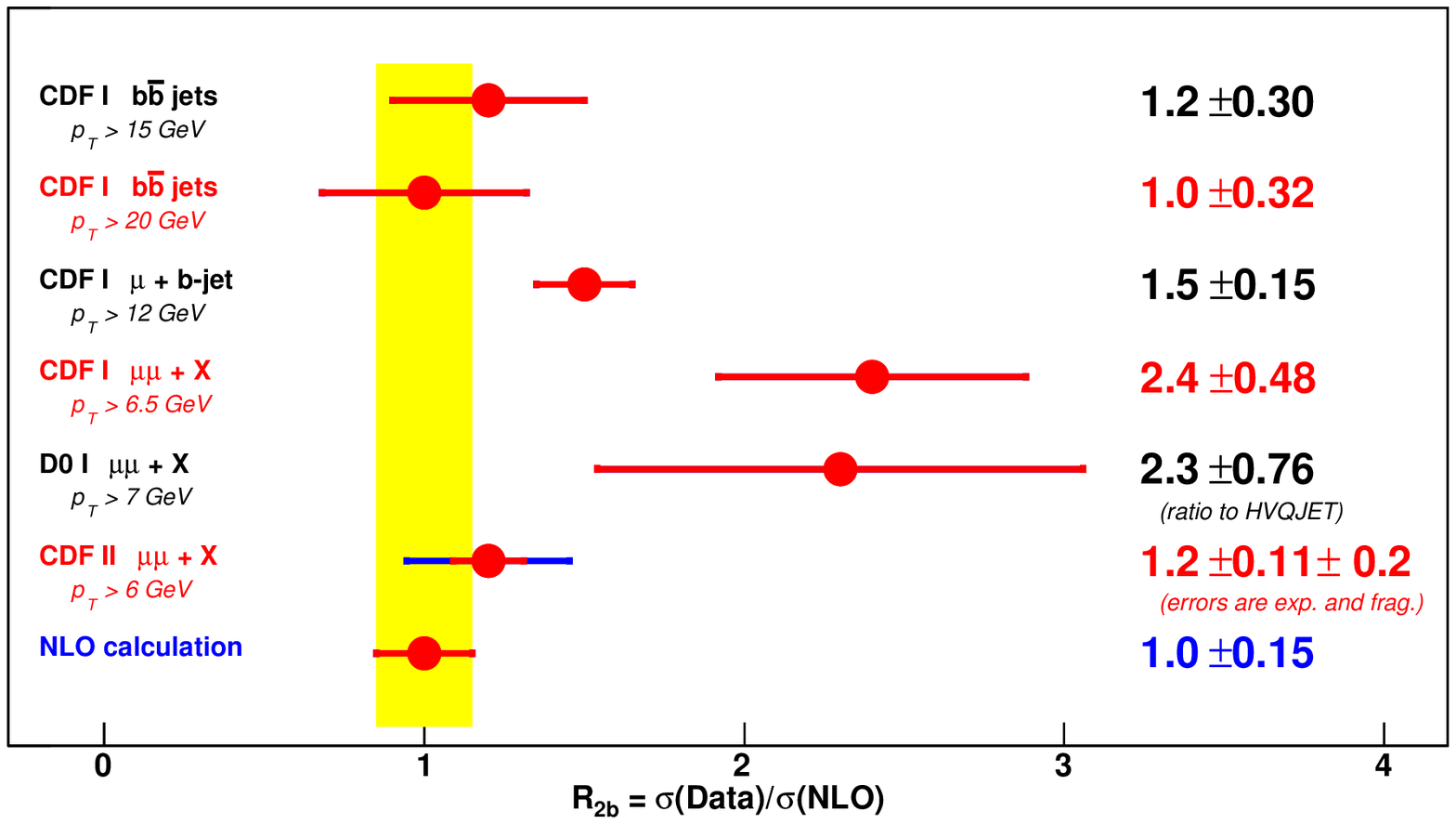}
\caption{\label{fig:comp2}Ratio of data/$\rm NLO_{theory}$ for correlated $b\bar{b}$ cross-section measurements.}
\end{minipage}
\end{figure}

The first result is a measurement of b-hadron ($H_b$) cross-section~\cite{Kraus:2006}.
This measurement uses $p\bar{p}\rightarrow H_b + X\rightarrow\mu^-D^0+X$ events collected with the lepton plus displaced track trigger introduced in Run~II.
A signal of ~3200 $\mu D^0$ candidates with $D^0\rightarrow K^- \pi^+$ is reconstructed in a sample corresponding to a luminosity of $83 \rm pb^{-1}$.
To obtain the cross-section, we first subtract the resonant background under the $D^0$ peak, which is estimated to be $12.0\%\pm3.2\%$ with a combination of data measurement and MC simulations.
Then we evaluate the acceptance with MC, while all the efficiencies are measured with data.
Finally we unfold the $p_T(\mu D^0)$ distribution using MC input to obtain 
the $H_b$ differential cross-section shown in figure~\ref{fig:comp1}.
The total cross-section is:
\[
 \sigma_{H_b} (p_T > 9.0 \, {\rm GeV/c}, |y|<0.6) =1.34 {\rm \mu b \pm 0.08 (stat.) \, ^{+0.13}_{-0.14} (syst.) \pm 0.07 (BR) }
\]

The second result is a measurement of $B^+$ production cross-section with fully reconstructed $p\bar{p}\rightarrow B^+ + X$, $B^+\rightarrow J/\psi K^+$ events~\cite{Abulencia:2006ps}. It is based on a sample of 8200 signal events corresponding to a luminosity of $740 \rm pb^{-1}$.
This analysis exploits the high-statistic available to simplify the selection and minimize the overall systematics.
The total cross-section is:
\[
 \sigma_{B^+} (p_T > 6.0 \, {\rm GeV/c}, |y|<1) =2.78\pm0.24 {\rm \mu b}
\]

The two new CDF~Run~II measurements are compared to the previous CDF~run~II measurement and with the lastest theoretical prediction~\cite{Cacciari:2003uh} in figure~\ref{fig:comp1}, where all results are scaled to $\sigma_{B^+}$. The agreement is quite good. It shows the result of several years of work to improve our understanding of bottom production.

\section{New measurement of correlated $b\bar{b}$ cross-section at CDF}

The review of $b\bar{b}$ cross-sections in ref.~\cite{Happacher:2005gx} shows that the five measurements from Run~I are not consistent among themselves. This is represented by the first five points in figure~\ref{fig:comp2}. 
It seems that the data/theory ratio increases with the numbers of muons in the observed final state.
A new measurement from CDF~\cite{CDFbbbar} addresses this anomalous effect.
It uses dimuon events selected with $p_T>3 {\rm \, GeV/c}$, $|\eta|<0.7$ and $5<m_{\mu\mu}<80 {\rm \, GeV/c^2}$. The invariant mass cut is used to reject events from sequential decays of a single b-quark and from the $Z^0$ resonance.
This dimuon sample includes events from different sources:
\begin{itemize}
\item decays of heavy flavor quark pairs ($b\bar{b}$, $c\bar{c}$)
\item prompt Drell Yan processes, charmonium and bottomoium
\item K and $\pi$ decays and misidentification
\end{itemize}
In order to separate the different contributions and eventually count the number of dimuon events from heavy flavor quark pairs we fit the 2D distribution of the impact parameter of both muons. We first generate templates of the 1D impact parameter distributions. For prompt single muons we use data and for single muons from b-hadron and c-hadron decays we use a tuned Herwig simulation including full CDF~II~detector simulation.
The three 1D templates obtained are combined into a six 2D templates for each possible dimuon source. Namely $b\bar{b}$, $c\bar{c}$, $cb$, prompt-prompt, prompt-$b$ and prompt-$c$.
The six templates are then used for a maximum likelihood fit of the measured 2D distribution.
We find $54583\pm678$ dimuons from $b\bar{b}$ and $24458\pm1565$ dimuons from $c\bar{c}$ in our sample that corresponds to a luminosity of $740 {\rm pb^{-1}}$.
The fraction of real muon pairs in our sample is estimated with both data and MC and it is found to be $0.96\pm0.04$ for $b\bar{b}$ and $0.81\pm0.09$ for $c\bar{c}$.
The overall acceptance times efficiency is estimated with MC and corrected with scale factors measured on data.

In table~\ref{tab:bbbar} we report our final results that are dimuon cross-sections from heavy flavor pairs instead of $b\bar{b}$/$c\bar{c}$ cross-sections.
Where $\sigma_{b\rightarrow \mu, \bar{b}\rightarrow \mu}$ is defined as 
$\sigma_{b\bar{b}} \cdot BR(b\rightarrow \mu +X)^2 \cdot A(p_T^\mu>3{\rm \, GeV/c},\, |\eta^\mu|<0.7,\, 5<m_{\mu\mu}<80 {\rm \, GeV/c^2})$.
The $\sigma_{b\rightarrow \mu, \bar{b}\rightarrow \mu}$ error is dominated by luminosity uncertainty. The $\sigma_{c\rightarrow \mu, \bar{c}\rightarrow \mu}$ error is dominated by systematics from the impact parameter fit and the fake muon removal.
In order to compare with theory, we calculate a prediction at dimuon level. We combine the MNR calculus with MRST98 PDF, Peterson fragmentation and we decay events with EvtGen as detailed in ref.~\cite{CDFbbbar}.
Table~\ref{tab:bbbar} reports the results for two different fragmentation parameters ($\epsilon$) in order to assess the uncertainty associated to fragmentation. 
Apart from fragmentation uncertainty, the theoretical error is 15\% from mass and scales systematics plus a BR uncertainty of 3.7\% and 6.7\% respectively for $b\bar{b}$ and $c\bar{c}$. 

\begin{table}[h]
\caption{\label{tab:bbbar}Data/theory comparisons for dimuon cross-sections from heavy flavor.}
\begin{center}
\begin{tabular}{llll}
\br
 & $\sigma_{b\rightarrow \mu, \bar{b}\rightarrow \mu}$ & $\sigma_{c\rightarrow \mu, \bar{c}\rightarrow \mu}$ \\
\mr
Data & $1549\pm133 \, {\rm pb}$ & $624\pm104 \, {\rm pb}$ \\ \hline
NLO & $1293 \, {\rm pb}$ ($\epsilon_b=0.006$) & $230 \, {\rm pb}$ ($\epsilon_c=0.06$) \\
NLO & $1543 \, {\rm pb}$ ($\epsilon_b=0.002$) & $383 \, {\rm pb}$ ($\epsilon_c=0.02$) \\
\br
\end{tabular}
\end{center}
\end{table}

The measured $\sigma_{c\rightarrow \mu, \bar{c}\rightarrow \mu}$ is higher than predictions.
This does not necessarily indicate a problem with $c\bar{c}$ cross-section.
In fact, the fragmentation uncertainty is quite large, and the result is just one and a half sigma higher than the theoretical prediction with $\epsilon_c=0.02$.
A more accurate comparison requires a dedicated tuning of fragmentation functions, e.g. following work from ref.~\cite{Cacciari:2005uk}.
The measured $\sigma_{b\rightarrow \mu, \bar{b}\rightarrow \mu}$ is compatible with theoretical predictions for both $\epsilon$ values. 
In order to compare with previous measurement, we convert $\sigma_{b\rightarrow \mu, \bar{b}\rightarrow \mu}$ into a measurement of $\sigma_{b\bar{b}}(p_T^b>6{\rm \, GeV/c},\, |y^b|<1)=1618\pm148\,(stat.+sys.)\pm312\,(frag.) \, {\rm nb}$ by scaling the MNR prediction for $\sigma_{b\bar{b}}$ using the data/theory ratio from table~\ref{tab:bbbar}. This provides the second last point in figure~\ref{fig:comp2}, that does {\it not} confirm the anomalously high dimuon cross-section seen in Run~I.

\section*{References}
\bibliography{iopart-num}

\end{document}